\documentclass[twocolumn]{aastex631} %
\usepackage[utf8]{inputenc}
\usepackage{hyperref}
\usepackage{float}
\usepackage{graphicx}
\usepackage{wrapfig}
\usepackage{tablefootnote}
\usepackage{threeparttable}
\usepackage{subfigure}
\usepackage{natbib}
\usepackage{hhline}
\usepackage{amsmath}
\usepackage{amssymb}

\begin{document}


\newcommand{\Cornell}{Department of Astronomy and Cornell Center for Astrophysics and Planetary Science, Cornell University, Ithaca, NY, 14853, USA}
\newcommand{\BLBerkeley}{Breakthrough Listen, University of California, Berkeley, 501 Campbell Hall \#3411, Berkeley, CA 94720, USA}
\newcommand{\UCB}{Department of Astronomy, University of California, Berkeley, 501 Campbell Hall \#3411, Berkeley, CA 94720, USA}
\newcommand{\seti}{SETI Institute, 339 Bernardo Ave, Suite 200, Mountain View, CA 94043, USA}
\newcommand{\KZA}{University of Malta, Institute of Space Sciences and Astronomy}
\newcommand{\GBO}{Green Bank Observatory, 155 Observatory Road, Green Bank, WV 24944, USA}
\newcommand{\Oxford}{Astrophysics, Department of Physics, University of Oxford, Denys Wilkinson Building, Keble Road, Oxford OX1 3RH, UK}
\newcommand{\BLOxford}{Breakthrough Listen, University of Oxford, Denys Wilkinson Building, Keble Road, Oxford OX1 3RH, UK}
\newcommand{\Manchester}{Department of Physics and Astronomy, University of Manchester, UK}

\newcommand{\tseti}{\texttt{turboSETI}}
\newcommand{\unit}{\dot{\nu}_{\text{unit}}}
\newcommand{\tsnr}{\ensuremath{\rho_{tseti}}}

\correspondingauthor{Ben Jacobson-Bell}
\email{bhj8@cornell.edu}

\author[0009-0009-6231-9280]{Ben Jacobson-Bell}
\affiliation{\BLBerkeley}
\affiliation{\Cornell}

\author[0000-0003-4823-129X]{Steve Croft}
\affiliation{\BLBerkeley}
\affiliation{\BLOxford}
\affiliation{\seti}

\author[0009-0008-0662-1293]{Carmen Choza}
\affiliation{\BLOxford}
\affiliation{\seti}

\author[0000-0003-2734-1895]{Alex Andersson}
\affiliation{\Oxford}

\author[0009-0007-3897-2912]{Daniel Bautista}
\affiliation{\GBO}

\author[0000-0002-8604-106X]{Vishal Gajjar}
\affiliation{\seti}
\affiliation{\UCB}

\author[0000-0002-7042-7566]{Matthew Lebofsky}
\affiliation{\UCB}

\author{David H.\ E.\ MacMahon}
\affiliation{\UCB}

\author[0009-0003-6274-657X]{Caleb Painter}
\affiliation{Department of Astronomy, Harvard University, Cambridge, MA 02138, USA}

\author[0000-0003-2828-7720]{Andrew P.\ V.\ Siemion}
\affiliation{\BLOxford}
\affiliation{\seti}
\affiliation{\UCB}
\affiliation{\KZA}
\affiliation{\Manchester}

\title{Anomaly Detection and Radio-frequency Interference Classification with Unsupervised Learning in Narrowband Radio Technosignature Searches}

\begin{abstract}
The search for radio technosignatures is an anomaly detection problem: Candidate signals represent needles of interest in the proverbial haystack of radio-frequency interference (RFI). Current search frameworks find an enormity of false-positive signals, especially in large surveys, requiring manual follow-up to a sometimes prohibitive degree. Unsupervised learning provides an algorithmic way to winnow the most anomalous signals from the chaff, as well as group together RFI signals that bear morphological similarities. We present GLOBULAR (Grouping Low-frequency Observations By Unsupervised Learning After Reduction) clustering, a signal processing method that uses HDBSCAN to reduce the false-positive rate and isolate outlier signals for further analysis. When combined with a standard narrowband signal detection and spatial filtering pipeline, such as \verb+turboSETI+, GLOBULAR clustering offers significant improvements in the false-positive rate over the standard pipeline alone, suggesting dramatic potential for the amelioration of manual follow-up requirements for future large surveys. By removing RFI signals in regions of high spectral occupancy, GLOBULAR clustering may also enable the detection of signals missed by the standard pipeline. We benchmark our method against the \citeauthor{Choza_2024} \verb+turboSETI+-only search of 97 nearby galaxies at $L$-band, demonstrating a false-positive hit reduction rate of 93.1\% and a false-positive event reduction rate of 99.3\%.

\end{abstract}


\section{Introduction}
Since the work of \citet{Cocconi:1959} and \citet{project_ozma}, radio frequencies have comprised the most widely explored domain in the search for extraterrestrial intelligence (SETI) due to their favorably low extinction across cosmic distances and the known efficacy of their use for communication by at least one spacefaring civilization---humanity. Since 2015, the Breakthrough Listen (BL) Initiative \citep{Worden:2017} has used various facilities, including the Robert C. Byrd Green Bank Telescope (GBT), to conduct radio-frequency searches of numerous targets, ranging in scale from the planetary \citep[e.g.,][]{Traas, Franz_2022} to the galactic \citep[e.g.,][]{Gajjar-2021, Choza_2024}, for unambiguously artificial signals, or ``technosignatures."  Such signals are generally assumed to be narrowband, with signal widths on the order of 1–10\,Hz, for transmission efficiency and since no known astrophysical process naturally produces emission so narrowly peaked. Continuous-wave emission is also typically assumed for feasibility of follow-up observations and while calculating transmitter distributions. Technosignature searches outside the radio have sought other types of signals, such as optical laser line emission \citep[e.g.,][]{Zuckerman_2023} and  infrared waste heat signatures \citep[e.g.,][]{Wright_2014, Hephaistos}. Additionally, some radio technosignature searches have relaxed the narrowband assumption or the continuous-wave assumption, such as, respectively, the searches by \citet{Gajjar-2022} for broadband pulsed signals with artificial dispersions and by \citet{Suresh_2023} for intermittent signals with artificial periodicities. However, these alternative premises necessitate different search methodologies, and only the narrowband, continuous-wave radio case is treated here.

The BL search for radio technosignatures, a project waged on petabytes of spectrogram data both archival and novel, is hindered by the enormous amount of radio-frequency interference (RFI) present at all observing bands. Since most RFI signals are themselves technosignatures, albeit with terrestrial origins, they are frequently very similar in morphology to the signals sought in BL searches. A robust filtration framework is therefore critical to distinguish desired signals from RFI.

Past GBT searches \citep[e.g.,][]{Enriquez, Price:2020} have primarily used two criteria to reject RFI. The first is that an extraterrestrial narrowband signal must exhibit a drift in frequency over time due to the relative motion between the transmitter and the telescope; any signal with a drift rate of 0\,Hz\,s$^{-1}$ is consequently rejected. The second is a spatial filter, motivated by the assumption that extraterrestrial signals should be localized on the sky whereas RFI often is not. Observations of a target are made in a cadence of three ``on" pointings interspersed with three ``off" pointings in an ABACAD pattern, where A designates an on-target scan and B, C, and D designate distinct off-target scans; any signal that appears in the ``offs" in addition to the ``ons" is likewise rejected. (For a full overview of the BL observation strategy, see Section 2.1 in \citet{lebofsky:19}.) Signals that pass these criteria are promoted from ``hits'' to ``events.''

However, despite the presence of these well-motivated filters in conventional search pipelines, false-positive events are still found in significant numbers, contributing to manual follow-up requirements that can be tedious for large surveys and may become prohibitive as data volumes continue to grow. Moreover, high RFI densities can confound the algorithms that match signals across pointings. For these reasons, RFI mitigation is as much a concern for technosignature searches as for any other area of radio astronomy. 

Past approaches in RFI mitigation have included software such as {\sc AOFlagger} \citep{AOFlagger}, which uses the \verb+SumThreshold+ method primarily to flag lines of RFI that are constant in either time or frequency \citep{Offringa_2010}. More diverse RFI morphologies have been sought by recent machine learning (ML) studies, which have included both supervised methods \citep[e.g.][]{Mesarcik_2023}, where the goal is to classify signals according to a known library of sources, and unsupervised methods \citep[e.g.,][]{Mesarcik_2022}, where the goal is to characterize patterns in a data set and identify outlier signals. However, most prior approaches do not generalize well to the unusually high spectral resolutions seen in BL data products \citep{lebofsky:19}, which are necessary for narrowband signal detection but reveal fine structures to which these approaches frequently are not sensitive. \citet{Pinchuk_2022} and \citet{Ma_2023} have demonstrated successful supervised methods for distinguishing technosignature candidates from RFI with a spatial filter taken into account. \citet{Choza_2024} also employed unsupervised learning in their use of DBSCAN to cluster their false-positive event set in a 2-dimensional feature space.

In this work, we present GLOBULAR (Grouping Low-frequency Observations By Unsupervised Learning After Rejection) clustering,\footnote{\url{https://github.com/bjacobell/gbt-hdbscan}} a method that leverages unsupervised learning to reduce false positives in technosignature searches before the spatial filter step by identifying common types of RFI at high spectral resolution and removing them from the data set. In past work, including \citet{Mesarcik_2022}, signals are not detected a priori, and so noise makes up the clusters while RFI signals are treated as outliers. By contrast, the GLOBULAR clustering pipeline builds off of existing detection algorithms. Once detections have already been made, common RFI types make up the clusters, and unusual RFI signals, as well as any potential extraterrestrial technosignatures, are the outliers.

\subsection{Unsupervised Learning}\label{UL}

Although RFI sources are diverse, many RFI signals bear significant morphological similarities, whether because they were produced by similar sources or by coincidence. For a set of $n$ statistical quantities, called ``features,'' that are sufficiently descriptive of a signal's structure, similarly shaped signals will occupy similar positions in the corresponding $n$-dimensional parameter space, called the ``feature space.'' Clustering algorithms, a form of unsupervised learning, present a way to quantify this proximity, grouping together signals that share similar structure.

Our approach implements the HDBSCAN (Hierarchical Density-Based Spatial Clustering of Applications with Noise) algorithm, first proposed by \citet{CMS_HDBSCAN} as an extension of DBSCAN \citep{DBSCAN}, and implemented for Python by \citet{McInnes2017}. HDBSCAN comes with a number of unique advantages over other clustering algorithms, including outlier detection and the ability to discover the number of clusters without user input. Additionally, unlike DBSCAN, which requires that the user prescribe a distance parameter $\epsilon$, HDBSCAN discovers clusters of different densities in a single run by varying over a preset range of values for $\epsilon$. However, a significant drawback, especially for radio astronomy applications, is that with a complexity $\mathcal{O}(N^2)$ that can be accelerated only up to $\mathcal{O}(N \log N)$ for certain data sets \citep{McInnes2017A}, HDBSCAN does not upscale easily to large data volumes. Previous methods of tackling the runtime problem include an implementation that uses the MapReduce framework across subsamples of the data set with a subsample size determined by single-unit processing power \citep{MR-HDBSCAN}. We employ the batching scheme described in Section \ref{batch}, which uses a subsample size determined by a preliminary analysis of the data, among other differences.

We apply HDBSCAN to the dual problem of RFI classification (grouping similar RFI signals together) and anomaly detection (isolating unusual signals for further analysis). Matching the classes discovered by our method to common sources of RFI, we demonstrate the validity of our clustering. We then apply the \verb+turboSETI+ \verb+FindEvent+ pipeline \citep{turboSETI} to the anomalous class and discuss the results.

\subsection{Data Sample}\label{data_sample}

To benchmark our method against an existing survey, we analyze the sample of nearby galaxies previously searched by \citet{Choza_2024}. The sample contains 97 of the 123 targets in the BL nearby galaxy sample \citep{isaacson:2017}, chosen because their declinations, all greater than $-20\degr$, make them observable with the GBT. We analyze only the $L$-band observations from this sample, a total of 100 cadences in the range from 1.1 to 1.9 GHz.

Although extragalactic technosignature searches require that any putative transmitters demonstrate truly prodigious power output to be detected, they constrain the existence of such transmitters by sampling trillions of star systems in a single survey (see discussion in \citeauthor{Choza_2024}). Extragalactic searches may also support unique RFI rejection methods, such as an extension of the scintillation filter used for a Galactic Center search by \citet{Brzycki_2024}, though these methods were not used by \citeauthor{Choza_2024} and are not further discussed in this paper.

With \tseti, \citet{Choza_2024} found 2,186,151 hits across the cadences in this band. However, we find that 268,084 of these hits are duplicates of other hits in the data set, a redundancy that does not affect the processing in the \verb+turboSETI+ pipeline,\footnote{Accordingly, the conclusions of \citeauthor{Choza_2024} are unaffected. The RFI spectral occupancy analysis by \citeauthor{Choza_2024} uses a different hit catalog that does not contain duplicates.} but does affect a density-based clustering algorithm. An additional 164 hits are too close to the edge of the band ($< 50$ kHz) for our statistical calculations to be effective (see Section \ref{features}, \#7 and \#8). We excise these hits from our analysis, leaving 1,917,903 unique hits for which our full feature list is calculable.

We also generate 100 synthetic events with \verb+setigen+ \citep{Brzycki_2022} with drift rates between $-3$ and $3$\,Hz\,s$^{-1}$ and SNRs between 1,000 and 10,000, corresponding to 300 hits since each event appears in three ``on'' pointings. These signals are designed to resemble the drifting, narrowband, spatially localized signals we aim to discover, modeled after observations of human-made spacecraft (e.g., Voyager 1) and expected to be similar for non-anthropogenic sources. By attempting to recover these signals, we estimate the algorithm's true-positive rate. We inject these events into the observation cadence for M\,31, found during our toy-model experimentation (Section \ref{toy}) to be an good example of a typically RFI-dense $L$-band observation, and run a \verb+turboSETI+ \verb+FindDoppler+ search to retrieve the hits. Only 296 of the 300 hits are recovered in this way for analysis in the next stage of the process, not an unexpected result as \verb+turboSETI+ routinely misses a small proportion of signals due to high drift rate, low power, intense RFI in nearby frequency channels, or other reasons. The four missed hits are in two pairs from two different events, so we have a theoretical recovery ceiling of 98 events.

For benchmarking, we use the \verb+turboSETI+ values found by \citet{Choza_2024} for all real hits, and an identical search to recover our injected hits. The searches were performed over a drift rate range from $-4$ to 4\,Hz\,s$^{-1}$, with a signal-to-noise threshold of 10 (see \citeauthor{Choza_2024} for a discussion of the limitations of \verb+turboSETI+'s detection algorithm and signal-to-noise calculations). 

\section{Feature Space}\label{features}

The features used in this work to represent signals are a set of prescriptive statistical parameters that were chosen due to their known relevance to the field. This is in contrast to the ``latent space'' of features produced by, for example, principal component analysis or the first half of an autoencoder \citep[e.g.,][]{Ma_2023}, which discover features algorithmically. The exact list was developed during preliminary analysis of a small portion of the data set (see Section \ref{toy}). A benefit to our approach is that the features have clear statistical and/or astrophysical significance to a human observer. While we generally do not know the features' relative importances --- i.e., the degree to which each feature influences the classification of a hit toward one cluster over another --- prior to setting them, we can estimate them a posteriori in various ways, including a random forest trained on the clustering results (Section \ref{random_forest}) and Shapley values (Section \ref{shap}). Our list is intended to be representative of the data rather than tailored specifically to any one algorithm, so while we prioritize HDBSCAN in this work, it may also provide a useful basis for methods such as isolation forests \cite{isolation_forest} or unsupervised random forests \cite[e.g.,][]{unsupervised_RF}.

Our feature space for this work is 13-dimensional, with 3 of the features calculated natively by the \verb+turboSETI+ implementation and the other 10 calculated afterward. Our feature list is effective, as we show through the feature importance analysis described above and in Sections \ref{random_forest} and \ref{shap}, but it is not exhaustive, and further work may refine the list by introducing new features or pruning relatively unimportant ones as necessary, especially for applications to other telescopes or observing bands.

We preprocess our calculated features to ensure that the scales are comparable from one feature to another. Features \#1 and 2 undergo quantile transforms, further discussed in their respective sections below. Features \#3, 5, 8, 10, 11, and 13 are converted to logarithmic scales. Features \#1, 3, 4, 5, 8, 10, 11, 12, and 13 are normalized to unit range, while Feature \#2 is normalized to unit variance following its quantile transform to a normal distribution and Feature \#9 is normalized to unit maximum with negative values permitted. Features \#6 and 7 do not undergo normalization because their ranges are already close to unity. The range of any feature can be adjusted to induce a corresponding effect on the feature's relative importance; we find, as shown in Sections \ref{random_forest} and \ref{shap}, that these ranges are effective for our clustering.

In greater detail, our feature space is as follows:

\begin{figure*}
    \centering
    \includegraphics[width=\textwidth]{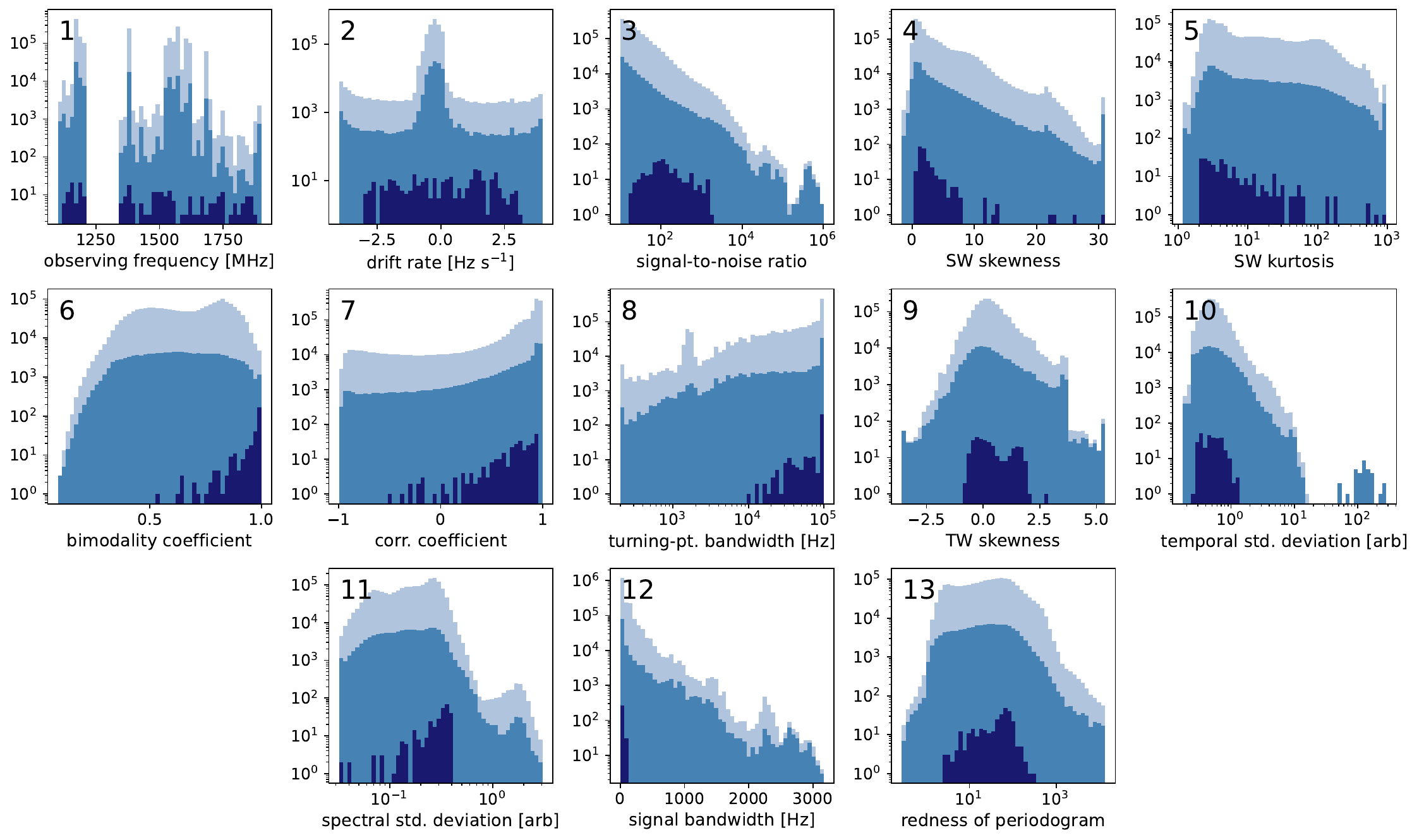}
    \caption{Distributions for the 13 features in our feature space, showing in light blue the distributions before any cluster-based rejection, in medium blue the distributions after 8 epochs of rejection, and in dark blue the retrieved distributions of the injected synthetic signals. Note the logarithmic scale on all $y$-axes. These distributions are from before the preprocessing steps outlined in Section \ref{features}, and therefore are not the distributions seen by the HDBSCAN algorithm, which are normalized to ranges of order unity (with the exceptions of Features \#6 and 7, which already have ranges of order unity). The gap in observing frequency between 1200 and 1340 MHz is due to a notch filter, the effects of which are removed in preprocessing.}
    \label{fig:1d_dists}
\end{figure*}

\textbf{\textit{\#1: Observation frequency.}} The observation frequency where a signal is detected is given by \verb+turboSETI+ in MHz. Observation frequency is an indirectly useful feature for clustering since morphologically similar RFI generally appears close together in frequency, due to allocations set by the Federal Communications Commission (FCC). However, frequency should not be considered at the expense of the morphological parameters; not all signals in the same band have the same source, and signals from similar sources can appear in noncontiguous bands. Frequency can be an unduly important feature for clustering because the spectral occupancy may be enormous for some bands, yet negligible for their neighbors, producing high-density clusters in the frequency dimension with relatively few outliers in between.

We preprocess the observation frequencies with a quantile transform to a uniform distribution, preserving the signals' relative locations in frequency space but eliminating those locations' relative densities. This preserves the relevant information, the apparent spectral proximity of morphologically similar signals to each other, while smoothing out the precise divisions set by the FCC  allocations.

\textbf{\textit{\#2: Drift rate.}} The drift rate is calculated by \verb+turboSETI+ and recorded in Hz\,s$^{-1}$. A high density of signals near zero drift will occur in virtually every data set due to the time-invariant nature of most RFI. If the density in this region is enormous compared to the wings of the distribution, these signals may be seen as a cluster and the relatively few signals in the wings as outliers, generally an incorrect interpretation. We preprocess the drift rates with a quantile transform to a Gaussian distribution to decrease the likelihood of this interpretation, aiding the algorithm's identification of low-drift anomalies at the expense of some high-drift sensitivity. This trade-off is desirable because, mainly due to the presence of high-drift transmitters (e.g., Iridium satellites) with a greater drift range than low-drift transmitters, the precise value of a drift rate is less important at high drifts than at low drifts.

We also take the magnitude of the drift rate, discarding the sign, again to improve clustering in the wings. This raises the likelihood of missing a technosignature in an environment of structurally similar RFI with exactly the opposite drift rate, but this could only occur in the first place by unlikely coincidence.

The drift rate distribution is discretized with a resolution $\Delta \dot{f} = \Delta \nu / T_\text{obs}$, where $\Delta \nu$ is the frequency resolution and $T_\text{obs}$ is the integration time of the data product used \citep{Siemion:2013}. For BL fine-frequency data products, $\Delta \dot{f} = 0.010204$\,Hz\,s$^{-1}$. The discretization may lead to inaccurate clustering due to the artificial gaps between populations, especially at drift rates near 0\,Hz\,s$^{-1}$, where the occupancy is high. Future work may benefit from mitigating the effect of these gaps.

\textbf{\textit{\#3: Signal-to-noise ratio (SNR).}} The SNR is calculated by \tseti. As \citet{Choza_2024} show, the \verb+turboSETI+ noise calculation typically underestimates SNR by a factor of $\sim 3.3$, such that a technosignature search with a prescribed SNR cutoff of 10 is in reality searching for signals above an SNR of $\sim 33$. The SNR distribution plotted in Figure \ref{fig:1d_dists} shows the raw \verb+turboSETI+ values without making this adjustment.

\textbf{\textit{\#4: Spectral window skewness (SWS).}} As reviewed by \citet{skew_kurt}, one measure for the skewness $\gamma$ of a sample of $n$ values, with a given value denoted $x_i$ with $i \in \{1,\hdots,n\}$, is
\begin{equation}
    \gamma = \frac{m_3}{m_2^{3/2}},
\end{equation}
where $m_j$ is the biased $j$th central moment, given by
\begin{equation}\label{moment}
    m_j = \frac{1}{n}\sum_{i=1}^n (x_i - \bar{x})^j.
\end{equation}
We calculate the SWS using the SciPy \verb+stats+ library \citep{SciPy} 
from a time-integrated spectrum of bandwidth 2.7\,kHz, following an analysis by \citet{Painter_2024} that found this bandwidth to adequately capture the spectral environment without drowning out the central signal. We use the SWS in tandem with the corresponding kurtosis (\#5) to quantify the occupancy of the signal in the 2.7-kHz window; high values for spectral skewness suggest that the signal is broad, highly drifted, and/or has a combed (multiple spectral lines) structure.

\textbf{\textit{\#5: Spectral window kurtosis (SWK).}} %
Similarly to the sample skewness (see \#4), the sample kurtosis $\kappa$ is calculated as \citep{skew_kurt}
\begin{equation}
    \kappa = \frac{m_4}{m_2^2},
\end{equation}
with the definition for $m_j$ given by Equation (\ref{moment}). For symmetry, we calculate the SWK from the same spectral window as the SWS. 

Like SWS, high values for SWK imply high spectral occupancy, suggesting broadness, high drift, and/or combedness. Note that SWK values can be large (up to $\sim$10$^3$) when the kurtic bandwidth (see \#7) is small. In general, we find empirically that the dependence of SWK on the bandwidth over which it's calculated is logarithmic, and that divergences from this behavior are evidence for interesting structures, including nearby signals. Features \#7 and \#8 quantify these structures.

Pearson's definition for kurtosis is used over the excess kurtosis (i.e., $\kappa_\text{exc} = \kappa - 3$, also called Fisher's definition). The calculation of Sarle's coefficient (\#6), to be constrained to unit range, requires the former.

\textbf{\textit{\#6: Spectral bimodality coefficient.}} The spectral bimodality coefficient is Sarle's bimodality coefficient applied to the intensity distribution of a power spectrum. It quantifies the bimodality of this distribution, which helps quantify the sharpness of the signal peak. The coefficient is calculated as \citep{sarle}
\begin{equation}
    b = \frac{\gamma^2 + 1}{\kappa},
\end{equation}
where $\gamma$ is the SWS (\#4) and $\kappa$ is the SWK (\#5). The bimodality coefficient is bounded between 0 and 1, with low values suggesting unimodality and high values suggesting, though not guaranteeing, bimodality. A uniform distribution has $b = 5/9$. 

Limitations of the bimodality coefficient in statistical inference are well documented, including that its probability density function is unknown and so statistical significance tests cannot be applied \citep[e.g.,][]{Knapp_2007, Pfister_2013}. The dip test of \citet{Hartigan_Hartigan} may provide a viable alternative for inference analysis using these data.

\textbf{\textit{\#7: Correlation coefficient between SWK and log(bandwidth).}} The kurtosis over a spectral window containing a signal is a function of the window's bandwidth (herein referred to as the kurtic bandwidth; i.e., the bandwidth over which kurtosis is calculated). Assuming a narrowband ($\sim$10-Hz) signal, at low kurtic bandwidths ($\sim 10^2 – 10^3$\,Hz) centered on the signal, kurtosis has a generally linear dependence on the log of the bandwidth. If there is an additional signal separated from the signal of interest by $\Delta\nu_0$, the kurtosis drops suddenly when the kurtic bandwidth equals $2\Delta\nu_0$ before continuing to increase linearly for bandwidths $> 2\Delta\nu_0$. Inevitably, a ``turning-point bandwidth" occurs for which the kurtosis is a maximum, after which kurtosis decreases with increasing bandwidth (see \#8). The limit is low kurtosis (3 for Pearson's definition), which corresponds to a vanishingly narrow signal in overwhelming Gaussian noise.

Because SWK is sensitive to spectral occupancy---that is, kurtosis is high for windows with many narrowband signals, including comb structures---the behavior of the kurtosis–log(bandwidth) function helps quantify the nearby environment. We use the Pearson correlation coefficient between kurtosis and log(bandwidth) as a proxy for this relationship's linearity. We obtain this value by calculating kurtosis for 50 bandwidths logarithmically spaced between 200\,Hz and 100\,kHz (also used for \#8), then using the bandwidths between 200\,Hz and 2.7\,kHz (see \#4) to calculate the correlation coefficient using the SciPy \verb+stats+ library.

\textbf{\textit{\#8: Bandwidth for which spectral kurtosis is a maximum.}} Also referred to herein as the turning-point bandwidth. See \#7 for motivation. The turning-point bandwidth is calculated using the same 50 kurtic bandwidths calculated for \#7. For better precision, instead of taking the sample with maximal kurtosis, we fit a parabola to this sample and the two adjacent samples and compute its vertex. We note that a significant number of our hits have turning-point bandwidths beyond the searched range, for which we make no provision except to set those values to the top of the range (100\,kHz). A future implementation of this algorithm could search higher kurtic bandwidths and potentially discover new outliers on the basis of anomalously high turning-point bandwidths, but the computational cost of more extensive searches swiftly becomes excessive.

\textbf{\textit{\#9: Temporal window skewness (TWS).}} The TWS is calculated similarly to the SWS (\#4), except on the full 5-min time series rather than a small chunk of the power spectrum. Indicates the duty cycle of the signal; high values suggest that the signal is pulsed or otherwise has a strong peak at some point in time. We note that the typical time series for our observations is extremely coarse, with only 16 samples in a $\sim 5$-minute observation. Future work will likely benefit from extracting temporal statistics from data products with finer time resolution \citep{lebofsky:19}.

\textbf{\textit{\#10: Time-series standard deviation.}} The standard deviation of the frequency-averaged time series is generally informative about temporal behavior, which is useful for similar reasons to the temporal skewness (\#9). We note, as for temporal skewness, that our time series are extremely coarse, and the precision of this parameter will likely benefit from calculation on data with higher time resolution.

\textbf{\textit{\#11: Power-spectrum standard deviation.}} The standard deviation of the time-averaged power spectrum is generally informative about spectral behavior.

\textbf{\textit{\#12: Signal bandwidth.}} We calculate the signal bandwidth using tools from \verb+blscint+ \citep{Brzycki_2023}, which records the width in bin units (converted to hertz in, e.g., Figure \ref{fig:1d_dists}). Signal bandwidth is defined here as the frequency interval at 1\% of a signal's maximum power. We search bandwidths up to 5\,kHz and find none above $\sim 3.1$\,kHz in our data set. The vast majority of signals are only a few hertz wide.

\textbf{\textit{\#13: Redness of spectral periodogram.}} To characterize comb-like spectral structures, we take the Lomb–Scargle periodogram \citep{Lomb, Scargle} of a 10-kHz window centered on each signal and calculate its ``redness''; that is, an estimate of the power contained in low-``frequency'' periodic signals.\footnote{We use scare quotes here to emphasize that the frequency and redness in question are not in the typical frequency domain---that is, the Fourier transform of the time domain, commonly expressed in cycles per second, or hertz---but rather in the Fourier transform of the frequency domain, expressed in cycles per hertz.} We consider the first fifth of the periodogram, which is approximately equivalent to the positive half of a Fourier transform since we are not missing any chunks of data in the frequency domain. From this section, we take the ratio of the average power in the first half to the average power in the second half, and use this value as a redness estimate. The Lomb–Scargle calculation is done using the Astropy implementation \citep{VanderPlas_2012, VanderPlas_2015}.

\section{Method}

\begin{figure*}
     \centering
     \includegraphics[width=0.8\linewidth]{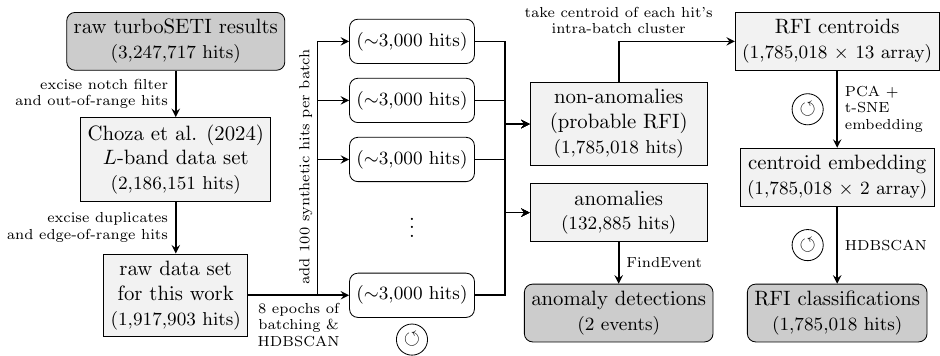}%
     \caption{End-to-end overview of the GLOBULAR clustering method. Numbers and some labels represent our implementation on the \citet{Choza_2024} data set and will vary by application. The raw data set also contains 296 synthetic hits corresponding to 98 events for true-positive recovery, which are not reflected in this chart. The three $\circlearrowleft$ indicate the iterations required to choose suitable hyperparameters for HDBSCAN and t-SNE; our heuristic criteria for making these choices are given in Section \ref{toy} and Section \ref{match}, respectively.}
     \label{fig:flowchart}
\end{figure*}

Due to the dimensionality of the problem in both feature count (13) and sample size ($\sim 2 \times 10^6$), the application of HDBSCAN to a large survey of narrowband radio signals is not trivial, requiring certain pre- and postprocessing measures to obtain reasonable results in both anomaly detection and RFI classification. A visual summary of these measures is given in Figure \ref{fig:flowchart}.

Since GLOBULAR clustering does not itself detect signals in spectrogram data, the method begins with a user-defined list of already discovered signals. We use those discovered by the \verb+turboSETI+ \verb+FindDoppler+ search algorithm on the GBT observations used by \citet{Choza_2024}, though in principle any list of suitably precise frequencies of detected signals will suffice. After excisions for the sensitivity range and duplicate signals, we have a data set of 1,917,903 hits (see Section \ref{data_sample}). We add the 296 synthetically generated hits recovered by \verb+FindDoppler+, bringing the pre-HDBSCAN data set to 1,918,199 hits. Only the \citeauthor{Choza_2024} hits, and not our injected ones, are reflected in Figure \ref{fig:flowchart}.

We calculate features for each hit (see Section \ref{features}), reducing the signal from a many-pixel array to a 13-number representation. After reduction, we pass the 13 features from all 1,918,199 hits in batches through successive instances of HDBSCAN, pooling and shuffling the batches with each iteration to ensure even clustering (see Section \ref{batch}). The iterative batching and pooling may be done an arbitrary number of times to achieve the desired reduction rate, though the likelihood of false negatives may increase if this number is too high. We run the algorithm for 8 iterations, termed ``epochs'' by analogy with training methods in supervised learning, to obtain a total reduction rate of 93.1\%. The epoch-by-epoch results are given in Table \ref{tab:epoch_table}.

The HDBSCAN iterations partition the data set into anomalies and non-anomalies (i.e., the most probable RFI), which are thenceforth treated separately. The anomalies are passed through the \verb+turboSETI+ \verb+FindEvent+ pipeline, which applies the spatial filter across pointings in each cadence. We follow standard BL procedure in rejecting signals that appear in some ``on'' pointings but not others, or appear in any of the ``off'' pointings. The resulting events are fewer in number and more likely to be truly unusual than the events found by a pipeline involving \verb+turboSETI+ alone. Importantly, in the cadences we pass to \verb+FindEvent+, we trim the clustered RFI signals from the ``on'' pointings but leave the ``off'' pointings unaltered. If the ``off'' pointings are likewise trimmed, our method risks finding new false positives by removing continuous signals in the ``offs'' without removing their counterparts in the ``ons,'' simulating a signal localized on the sky. This produces approximately 50 new false-positive events in our testing. However, a case can be made for trimming the ``off'' pointings in spite of these new false positives. A sufficiently dense RFI environment may cause a signal to seem to be in one or more ``offs'' due to the presence of nearby signals (or superpositions of signals) with similar drift rates, even though their structure is different. GLOBULAR clustering could de-densify such an environment and remove the confounders, enabling an event detection that would not have been possible with untrimmed ``offs.''

The non-anomalies are clustered in each batch, but still need to be matched to superclusters across batches to characterize the full data set. We do this by first using principal component analysis (PCA) to project the intra-batch cluster centroids into a 6-dimensional space, then using $t$-distributed stochastic neighbor embedding (t-SNE) to project them into a 2-dimensional space, and then using HDBSCAN to cluster them. 

The two dimensions of the embedding contain the information of the input 6-dimensional space as the result of a non-linear mapping, and so have no intuitive physical meaning in either unit or scale. A benefit of t-SNE is that by tuning hyperparameters, the user can place varying degrees of emphasis on local or global structure, seeking a middle ground that emphasizes divisions between groups of clusters and aids matching. However, the ideal tuning is subjective, and, combined with the difficulty of interpreting apparent structures in t-SNE embeddings \citep{wattenberg2016how}, can easily lead to faulty results. We emphasize that visually spot-checking the clusters found by HDBSCAN at this stage is valuable for verifying that the output is sensible. We also note that empirically, we find substantial improvements in matching when we employ the dimensionality reduction measures discussed above, including improvements in the embedding when it is calculated from a reduced 6-dimensional space instead of the original 13-dimensional space. 

The following subsections outline the batching, iteration, and cross-batch matching processes in more detail.

\begin{center}
\begin{table}
\begin{threeparttable}
\vspace{0.6cm}
    \caption{Results of epochal iteration of HDBSCAN.}
    \begin{tabular}{c c c c}
        \hline
        \hline
              &           & Pct. reduction & Pct. of \\
        Epoch & Anomalies & from previous  & original \\
              &           & epoch          & data set \\
        \hline
        0\tnote{a} & 1,918,199 & --- & 100.0\% \\
        1 & 1,005,473 & 47.6\% & 52.4\% \\
        2 & 749,904 & 25.4\% & 39.1\%  \\
        3 & 585,257 & 22.0\% & 30.5\%  \\
        4 & 452,430 & 22.7\% & 23.6\%  \\
        5 & 330,067 & 27.0\% & 17.2\%  \\
        6 & 250,142 & 24.2\% & 13.0\%  \\
        7 & 190,122 & 24.0\% & 9.9\%  \\
        8 & 133,149 & 30.0\% & 6.9\%  \\
        \hline
    \end{tabular}
    \begin{tablenotes}
        \footnotesize
        \item[a] i.e., before any iterations.
    \end{tablenotes}
    \label{tab:epoch_table}
\end{threeparttable}
\end{table}
\end{center}

\subsection{Batching and Epochal Iteration}\label{batch}

In our testing, HDBSCAN lost sensitivity at high data volumes, having a tendency to group an extreme majority ($>$90\%) of hits in one cluster. We combat this problem by pre-batching our data and running HDBSCAN on a set of only $\sim$3,000 hits at a time, randomly sampled from the preprocessed ensemble. For our data set, this results in 639 batches. This approach risks splitting up small clusters into pieces too small to be caught by the algorithm, so we iteratively pool the anomalies from each batch into a new data set and run the algorithm again for further reduction. 

We deployed our method on the $L$-band subset of the observations used by \citet{Choza_2024}, a total of 2,186,151 hits (1,917,903 after our excisions and 1,918,199 after our injections; see Section \ref{data_sample}). 
The full data set is preprocessed before batching to ensure comparable ranges for all features, and batches are drawn at random from the ensemble with each hit having an equal probability of being drawn. Preprocessing helps ensure that all features are scaled proportionally and that the scaling is not unique on a per-batch basis. 

To each batch, we inject an identical set of $100 \times 13$ features derived from 100 drifting narrowband signals synthesized with \verb+setigen+. These signals are generated in the same way as, but are otherwise distinct from, the 100 events injected for recovery; the purpose of these is to ``seed'' clusters with morphologies we want to keep. Real drifting narrowband signals in the data set are likely to be clustered with the seed signals, enabling the user to more effectively track them through the epochal iteration. Any non-synthetic signals placed in seeded clusters are retained for the next epoch as though they were outliers.

\begin{figure}
    \centering
    \includegraphics[width=\columnwidth]{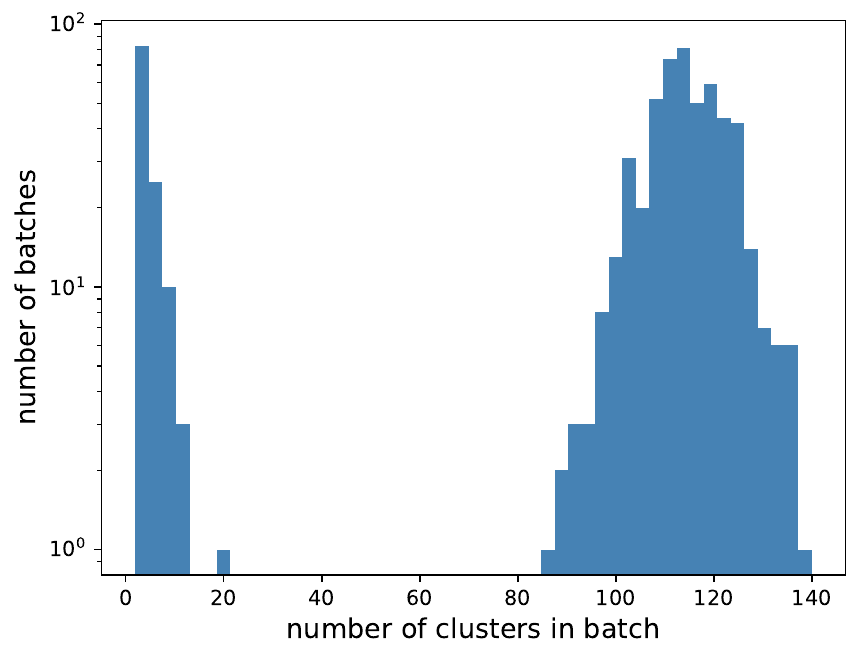}
    \caption{Cluster counts for 639 batches, each with $n \approx$ 3,000. 80.9\% of the batches are in the right-hand population and incur data reduction by discarding the non-anomalous classes. The batches in the left-hand population are insensitively clustered and do not incur reduction before the next HDBSCAN iteration.}
    \label{fig:hyp}
\end{figure}

The \verb+scikit-learn+ implementation of HDBSCAN, which we used for our analysis, takes an assortment of hyperparameters as user-defined inputs. We varied three: the minimum number of points per cluster $n_\text{pts}$; the minimum number of neighborhood points for clustering (a proxy for cluster density) $\rho_\text{pts}$; and the merging threshold $\epsilon_m$,\footnote{Not to be confused with the DBSCAN hyperparameter $\epsilon$ (see Section \ref{UL}).} which is a distance parameter proposed by \citet{Malzer_HDBSCANe} that prescribes the merging of clusters closer together in feature space than its value. We discuss our methodology for tuning these hyperparameters in Section \ref{toy}.
Figure \ref{fig:hyp} shows the cluster count per batch for the first epoch, with $n_\text{pts} = 4$, $\rho_\text{pts} = 2$, and $\epsilon_m = 0.18$. Two batch populations are evident: the one on the left, with very small cluster counts; and the one on the right, with cluster counts in the $\sim$10$^2$ range. The left-hand population corresponds to batches that are clustered insensitively; that is, an extreme majority of hits ($\gtrsim$ 90\%) are grouped in a single cluster. The right-hand population corresponds to batches with more sensitive, and accurate, clustering.

An additional benefit to batching for large data sets is that the complexity of the HDBSCAN algorithm is generally $\mathcal{O}(N^2)$, so compute time rises quadratically with sample size $N$ for an unbatched data set. However, suppose the data set is split into $k$ batches, each of size $N/k$. Suppose also that the compute time for the unbatched data set is $A N^2$ for some arbitrary coefficient $A$. The compute time for all batches is $k \times A(N/k)^2 = AN^2/k$, a factor of $k$ speed-up from the unbatched case. Furthermore, batches can be processed in parallel, permitting total acceleration by a factor of up to $k^2$. The compute time for our unbatched data set was $\sim 13$ hr, whereas by batching with $k= 639$, we processed all batches for the first epoch serially in just a few minutes.

We ran HDBSCAN for 8 epochs, obtaining a total hit reduction rate of 93.1\%. The corresponding event reduction rate, when \verb+FindEvent+ is deployed on the reduced hit set, is 99.3\% (see Section \ref{results}). The epoch-by-epoch results are given in Table \ref{tab:epoch_table}.

\subsection{Cross-Batch Cluster Matching}\label{match}

\begin{figure}
    \centering
    \includegraphics[width=\columnwidth]{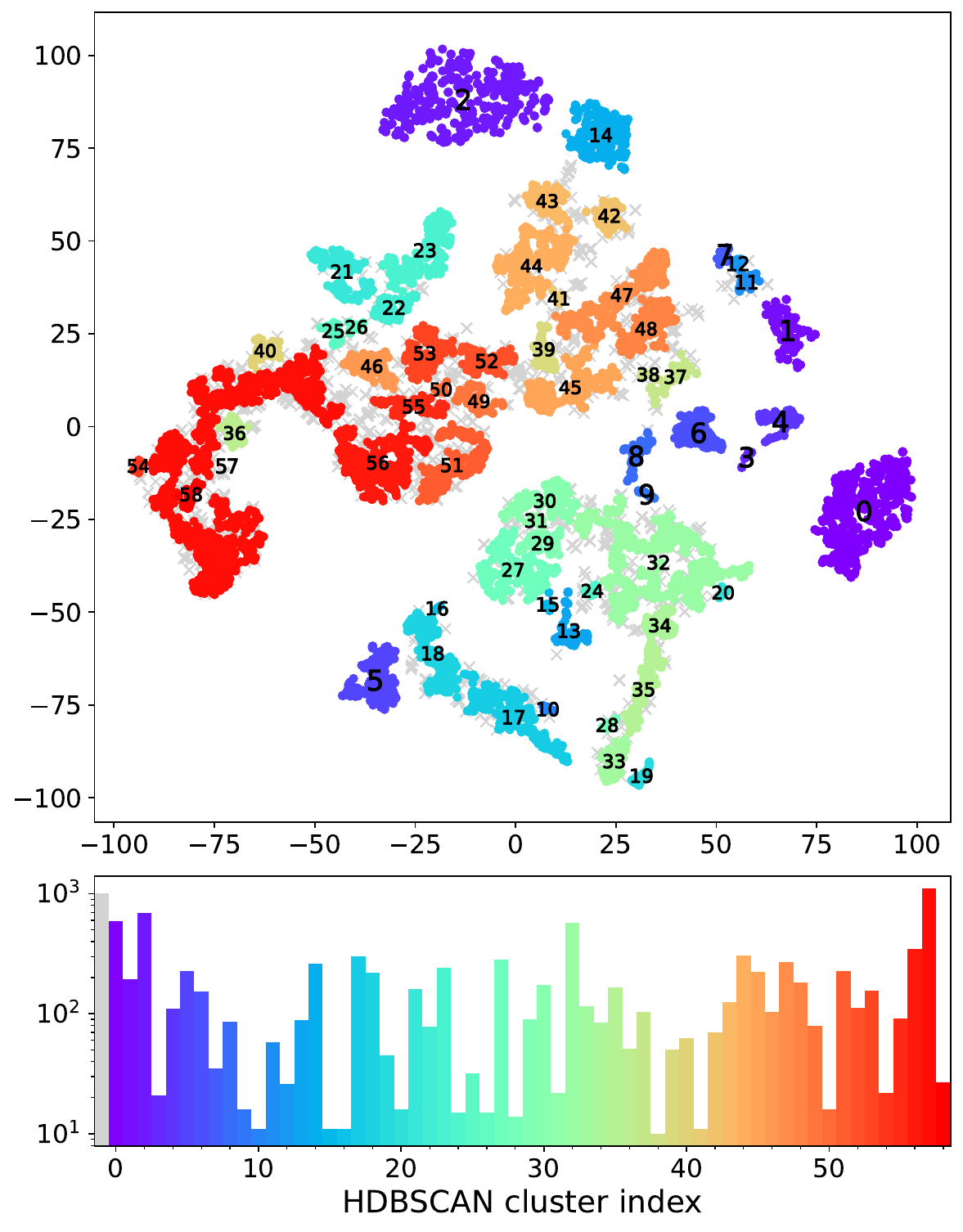}
    \caption{\emph{Top:} 2-dimensional t-SNE embedding of a 10,000-hit sample. The axes are derived from a non-linear mapping and have no intuitive physical meaning. Each hit is passed through t-SNE not with its actual position in our feature space, but with the centroid of the cluster it was found in during the 8 epochs of batching. We run HDBSCAN on the t-SNE embedding to identify which centroids correspond to morphologically similar clusters. HDBSCAN clusters are color-coded by index, with hits in the outlier cluster (indexed $-1$) represented by light gray $\times$'s. The 59 clusters shown here are an estimate, not a 1-to-1 identification, of common RFI sources in the environment. \emph{Bottom:} Histogram showing hit counts for each cluster.}
    \label{fig:tSNE}
\end{figure}

Since the number of clusters found by HDBSCAN may vary from one batch to the next (see Figure \ref{fig:hyp}), characterizing the RFI environment of a large data set requires that similar clusters be identified across batches. We do this by matching cluster centroids, assuming that any two clusters with proximate centroids in our feature space are likely to contain RFI with similar structure.

The t-SNE algorithm offers one way to improve the matching. We key each hit to the centroid of the cluster it was found in during the epochal iteration. This means that, for a cluster with 100 hits, its centroid will be repeated in the pre-embedding data set 100 times. We run a combination of PCA and t-SNE on random samples of 10,000 centroids, where repetition in the sample is possible due to repetition in the ensemble. We employ the first 6 principal components for data reduction before t-SNE, finding that this is the number necessary to capture more than 95\% of the variance in the centroid set. The result is a 2-dimensional embedding like the one shown in Figure \ref{fig:tSNE}. Distinct groups of centroids are made evident by the embedding, so we run HDBSCAN on the embedded points to recover these groups. The colors in Figure \ref{fig:tSNE} show the HDBSCAN labels.

The main hyperparameter needing to be tuned in the t-SNE algorithm is the perplexity, which indicates how much local versus global structure is emphasized in the embedding. Higher perplexity values tend to mean more clearly defined separations between higher-level clusters at the expense of low-level detail. We find that perplexities between 15 and 40 work well for our data, and that the algorithm is fairly but not entirely robust to variation within this range. Varying the size of the sample does appear to affect the appropriate perplexity range. We also tune the early exaggeration, a hyperparameter that helps emphasize the separation between clusters, which can aid HDBSCAN's identification of them. We use early exaggeration values of order 1–10. The embedding in Figure \ref{fig:tSNE} was produced with perplexity 40 and early exaggeration 4.

Since distances in a t-SNE embedding are the result of a non-linear mapping and do not necessarily have an intuitive distance-like meaning, apparent clusters in the embedding must be interpreted with caution. Relatively low-level structure in the embedding may in fact be important, yet may be ignored by HDBSCAN in favor of high-level clusters. Conversely, HDBSCAN may find several small clusters where a single supercluster would be more appropriate. There is a trade-off between cluster homogeneity and cluster redundancy. With the embedding in Figure \ref{fig:tSNE}, we have chosen to permit some redundancy in the interest of greater homogeneity.

A hierarchical approach may improve these results, choosing input hyperparameters that permit some inhomogeneity at first, and then requiring a re-embedding and re-clustering of each cluster that appears inhomogeneous. Another option is active learning, which can leverage user input to slightly adjust clusters for greater homogeneity. We defer these improvements to future work, emphasizing for now that this method produces good clusters in general, even for large data sets, and that refinement is still possible.

\section{Application to Small Data Sets}\label{toy}

We developed our feature space in iterations of the HDBSCAN algorithm on a toy data set of 3,068 hits discovered with \verb+turboSETI+ in a 5-minute observation of the galaxy M\,31, over 187.5\,MHz of frequency space from 1501.5 to 1689.0\,MHz. 
We took 316 hits from a small frequency range around 1620\,MHz of a separate observation, of the star HIP\,3223, to reinforce the sparse population of a type of RFI signal we consider significant. Additionally, 3 artificial injections were made to the M\,31 observation using \verb+setigen+ and their parameters retrieved; these injections were tracked to ensure our toy-model clustering remained sensitive to likely technosignature candidates. The 187.5-MHz M\,31 observation used to test the toy model was a portion of the 800-MHz M\,31 observation to which we later injected 100 synthetic events for large-scale retrieval testing (Section \ref{data_sample}), and the 3 injections made for the toy model were generated using the same parameters as the 100 events for the large-scale model.

On our toy data set, with the right choice of hyperparameters, we achieved remarkably accurate clustering of common types of RFI signals at the same time as a reduction rate greater than 90\%. The best hyperparameters we found for this data set were $n_\text{pts} = 7$, $\rho_\text{pts} = 2$, and $\epsilon_m = 0.24$, and were chosen by the following process, which evolved from experimentation and empirical findings: 

\begin{itemize}

\item $n_\text{pts}$ is chosen first. With $\rho_\text{pts}$ and $\epsilon_m$ set to their minimum values (2 and 0, respectively), $n_\text{pts}$ is varied over the integers from 3 to approximately 10. We find that there is generally a threshold above which the clustering is unstable and a clear majority of hits (at least 60\%, but sometimes even above 90\%) are erroneously placed in one cluster. It is desirable to set $n_\text{pts}$ as high as possible without breaching this stability threshold, which takes some iteration.

\item $\rho_\text{pts}$ is set to 2. Due to the diffuseness in feature space of our data set, so long as $n_\text{pts}$ is set below the stability threshold, clustering is seldom helped by setting $\rho_\text{pts} > 2$. Increasing $\rho_\text{pts}$ has the effect of making the clusters more conservative, giving the user a higher degree of confidence that their clusters represent homogeneous sources of RFI. We are already confident in this due to our liberal choice of $n_\text{pts}$, so the main effect of $\rho_\text{pts}$ is to peel away signals from clusters and increase the size of the anomalous class. Instead, our method is to discover many prospective subclusters with a low $n_\text{pts}$ and $\rho_\text{pts}$, and then merge them with a well-chosen $\epsilon_m$.

\item Once $n_\text{pts}$ is set, $\epsilon_m$ is varied between approximately 0.15 and 0.25 in increments of 0.01. Increments finer than this can be meaningful on small-scale data sets for which the goal is to refine the clusters as precisely as possible, but if the data set is large enough to warrant batching (see Section \ref{batch}), lower precision is acceptable. Here, as with $n_\text{pts}$, we aim to set $\epsilon_m$ just slightly lower than a certain stability threshold.

\end{itemize}

If the analysis is performed on only a single batch's worth of signals, the computational cost of iterating many runs to optimize these hyperparameters is low. However, in the case that significantly broader ranges of values need to be tested, or that the batch size is significantly larger than we use here, utilities such as CORE-SG \citep{CORE-SG} and MustaCHE \citep{MustaCHE} may be useful for efficiently iterating over $n_\text{pts}$ values and for visualizing the output, respectively.

The advantage of HDBSCAN over DBSCAN---that HDBSCAN is hierarchical and thus invariant with respect to cluster density---is mitigated somewhat by the use of a non-zero $\epsilon_m$, which introduces a hyperparameter that functions differently for subclusters of different densities. However, this effect also enables the productive use of consecutive HDBSCANs to cluster anomalies more precisely. For example, when upscaling our toy model to include the hits from a second observation
, approximately doubling our data set's size, we could only obtain a reduction rate of $\sim$70\% with a single HDBSCAN, but by isolating the anomalies from the first run and processing them again with a second, differently-tuned HDBSCAN, we increased the hit reduction rate to $\sim$95\%. Consecutive HDBSCANs are similarly useful for implementations of this algorithm on large data sets, as discussed in Section \ref{batch}.

\subsection{Random Forest Analysis}\label{random_forest}

\begin{figure}
    \centering
    \includegraphics[width=\columnwidth]{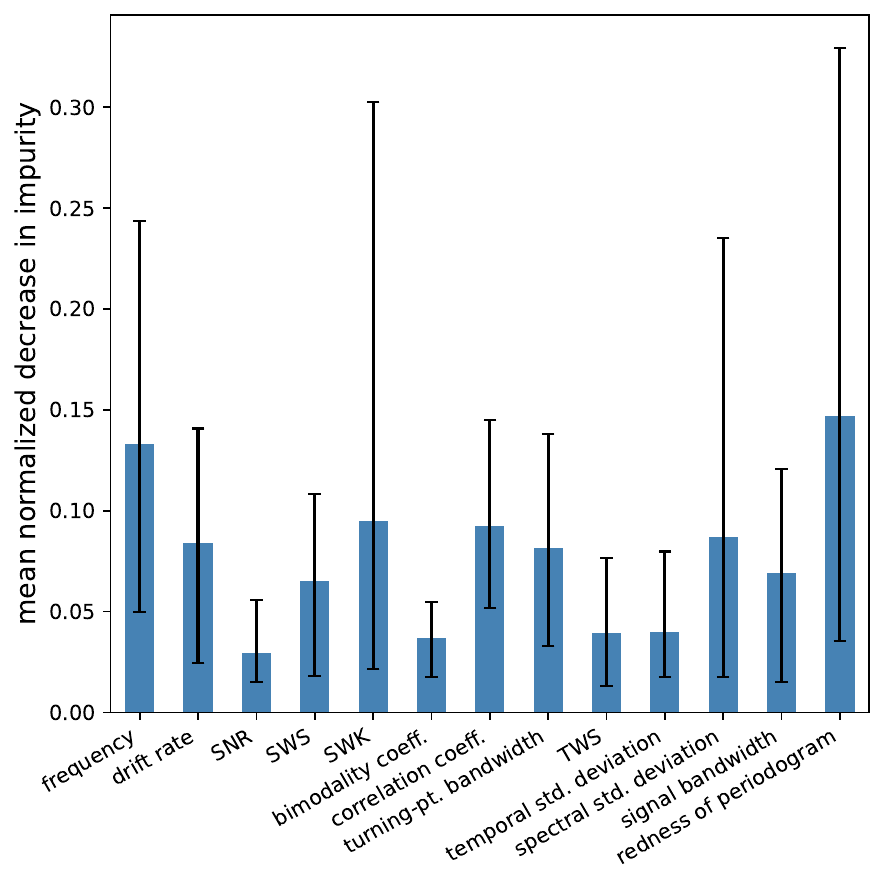}
    \caption{Feature importances from a random forest trained on a 3,387-signal toy data set. Importances are calculated by normalizing the decrease in Gini impurity for each tree to unit sum, then averaging across all 128 trees. Error bars contain 68\% of trees.}
    \label{fig:importances}
\end{figure}

An instance of supervised learning assessing the validity of unsupervised learning, random forests \citep{BreimanRF} offer a rough yet effective way to estimate the relative importances of the features in our feature space. The idea centers on the fact that each node in each tree of a random forest---the point at which a decision is made about whether to classify an input signal one way or another---uses one of the input features to make its decision. This is why it is generally much more effective to use a random forest to classify from a low-dimensional latent space, as in \citet{Ma_2023}'s use with a $\beta$-variational autoencoder, than from an uncompressed image with many unimportant elements.

In training the random forest, a loss function---frequently the Gini impurity; see \citet{Lin_2024} for an overview---is minimized, and each node has a quantifiable contribution to the minimization. By keeping track of the decrease in loss due to each node and averaging over all the nodes in the entire forest that correspond to a single feature, we can use the feature's mean decrease in impurity as a proxy for its importance. The \verb+scikit-learn+ random forest implementation has an in-built way to track these values.

Figure \ref{fig:importances} shows the mean decrease in Gini impurity for each feature when the toy data set of 3,387 points is used to train a random forest. The forest consists of 128 decision trees, each of which sees all 13 features.
Labels were assigned by HDBSCAN clustering.

The error bars encompass 68\% of the trees in the forest, and show quite a large spread for some features. The error bars on feature importances typically decrease for larger data sets; it is rarely good practice to train any supervised ML model on a data set of only a few thousand samples. However, a larger data set presents another challenge in that batching (see below) does not guarantee that morphologically similar clusters are indexed the same way across the entire data set. Therefore, when upscaling this approach, careful accounting is needed to match clusters across batches.

The benefit of this approach comes from demonstrating in a general sense that no one feature is significantly more important than the others. If one were, it could cause the HDBSCAN algorithm to cluster on the basis of it alone. If a feature does turn out to be unusually important, measures can be taken to reduce its importance, such as downscaling its range in preprocessing.

\subsection{Shapley Values}\label{shap}

\begin{figure}
    \centering
    \includegraphics[width=\columnwidth]{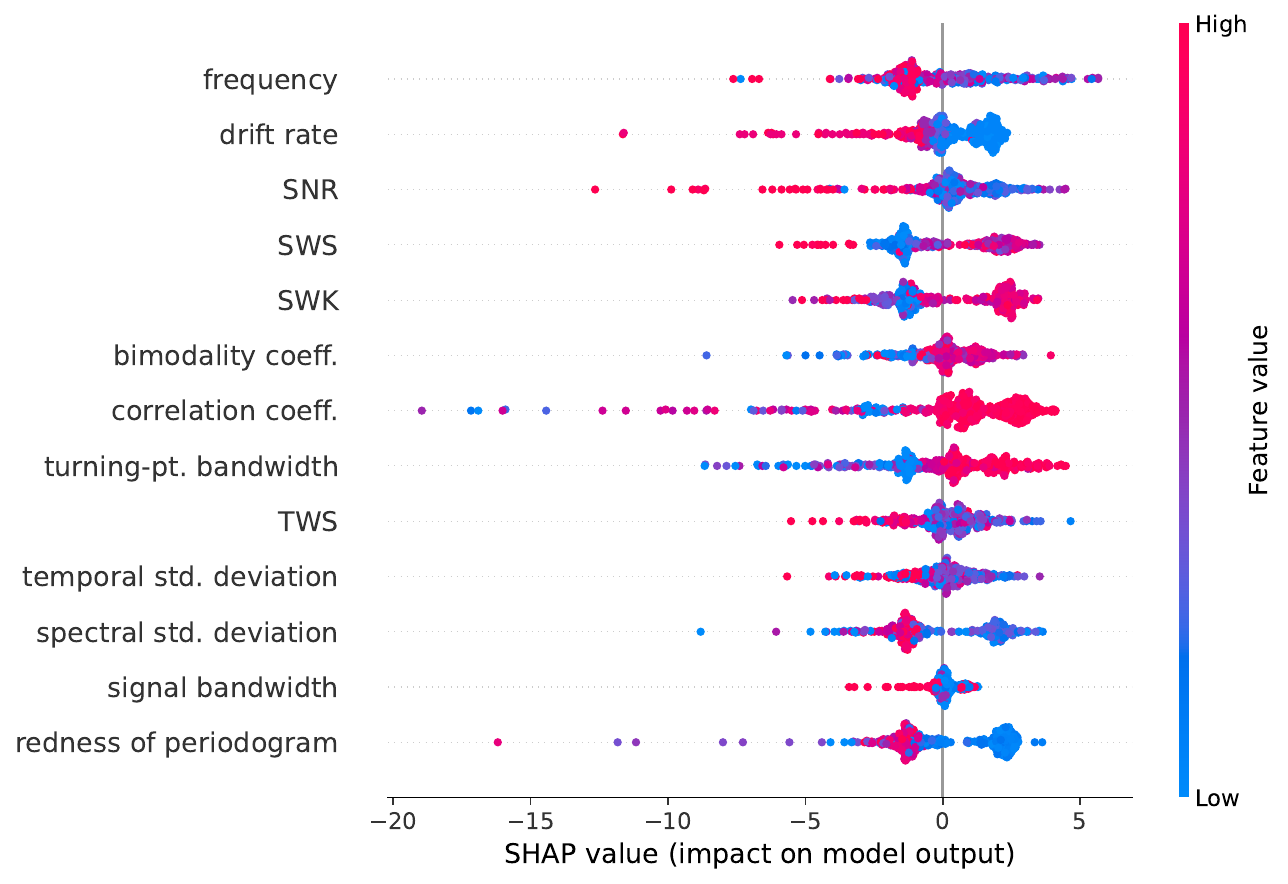}
    \caption{SHAP beeswarm plot generated from an HDBSCAN model's cluster predictions of 400 hits. The model was trained on the same 3,387-hit toy data set as the random forest in Section \ref{random_forest}. Each point represents a single feature for a single hit, and its color corresponds to the feature's value. For example, the pinkest point along the SNR axis corresponds to the hit with the highest SNR.} 
    \label{fig:shap}
\end{figure}

Shapley values, from game theory \citep{shapley1951notes}, have been used in astronomical ML analysis \citep[e.g.,][]{Gilda_2022, Reddy_2023} to quantify and allocate credit for a model's optimization among the features that affect it. For unsupervised learning, this is done by having a model that has already clustered a data set predict the clusters of a small sample of new points. For this, we use the \verb+approximate_predict+ method of the HDBSCAN implementation by \citet{McInnes2017}, which does not have an equivalent in the \verb+scikit-learn+ implementation as of version 1.5.2.

We calculate Shapley values using SHapley Additive exPlanations (SHAP), a package by \citet{shap}. Using an HDBSCAN model trained on the same 3,387-hit data set as the random forest in Section \ref{random_forest}, we use a \verb+KernelExplainer+ with a background sample of 100 hits to explain the predictions of 400 hits, drawn at random from the toy data set. We obtain the beeswarm plot in Figure \ref{fig:shap}, which shows that the features' Shapley values are generally well distributed according to their actual values. That is, for example, high-SNR hits affect the model differently than low-SNR hits, as they should. In all cases, the mean of the absolute value of all Shapley values for a feature is significantly above zero, but comparable across features to within an order of magnitude. This suggests, like Figure \ref{fig:importances}, that the features have similar importances on average. 

The features derived from the frequency-averaged time series, the temporal skewness and the temporal standard deviation, are significantly noisier than other features, with hits of many different colors clumping together around the mean. This aligns with our expectations; since the time series is only 16 bins of $\sim$18 s each, any values we derive from it are bound to be noisy. As we suggest in Section \ref{features}, future work can improve these features by calculating them from data products with higher time resolution.

\subsection{Principal Component Analysis}

\begin{figure}
    \centering
    \includegraphics[width=\columnwidth]{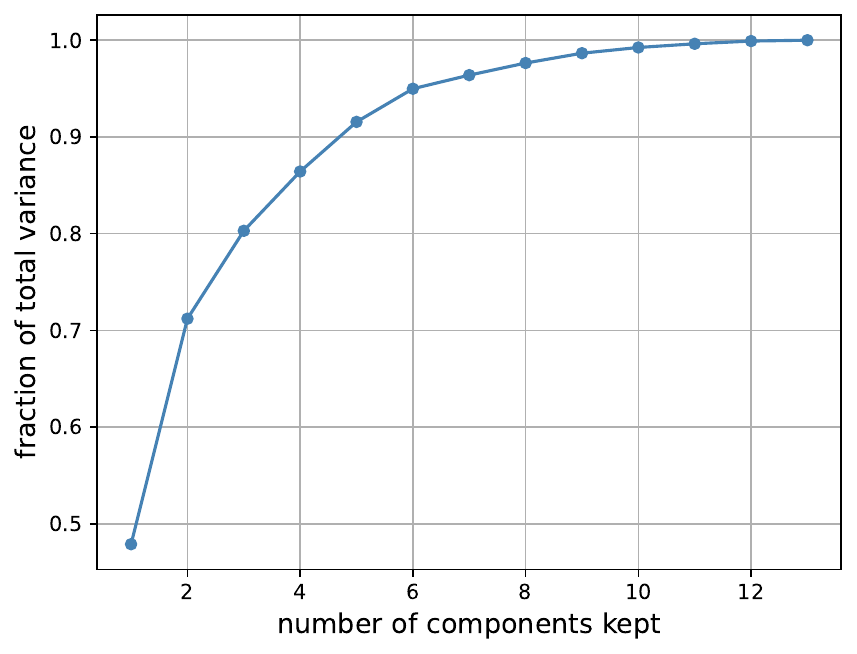}
    \caption{Fraction of total variance explained by keeping different numbers of principal components. 
100\% is only attained by keeping all 13, but $>$99\% can be attained with only 10. We note that even 1\% of the variance is significant for anomaly detection, and keep all 13 original features in our analysis.}
    \label{fig:PCA}
\end{figure}

\begin{figure*}
    \centering
    \includegraphics[width=0.49\textwidth]{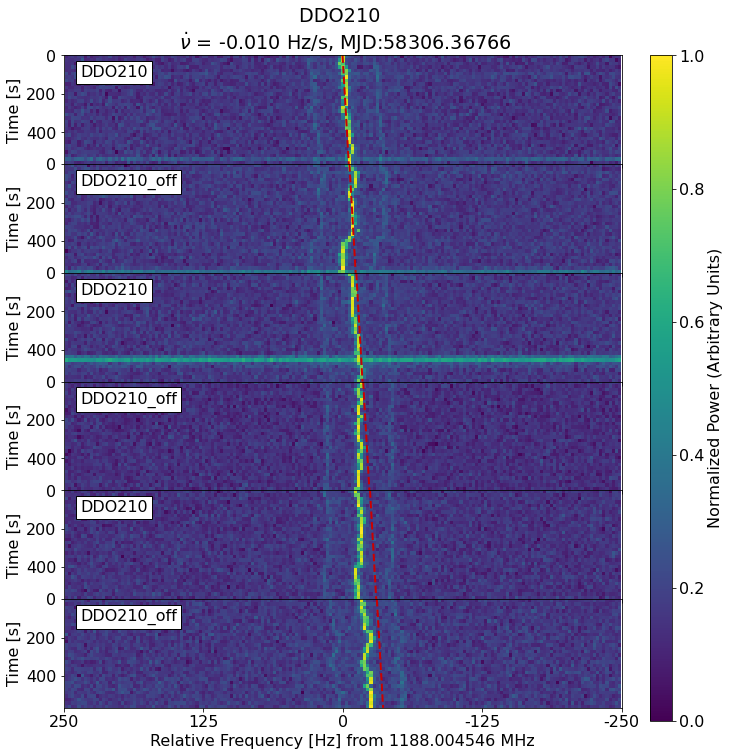}
    \includegraphics[width=0.49\textwidth]{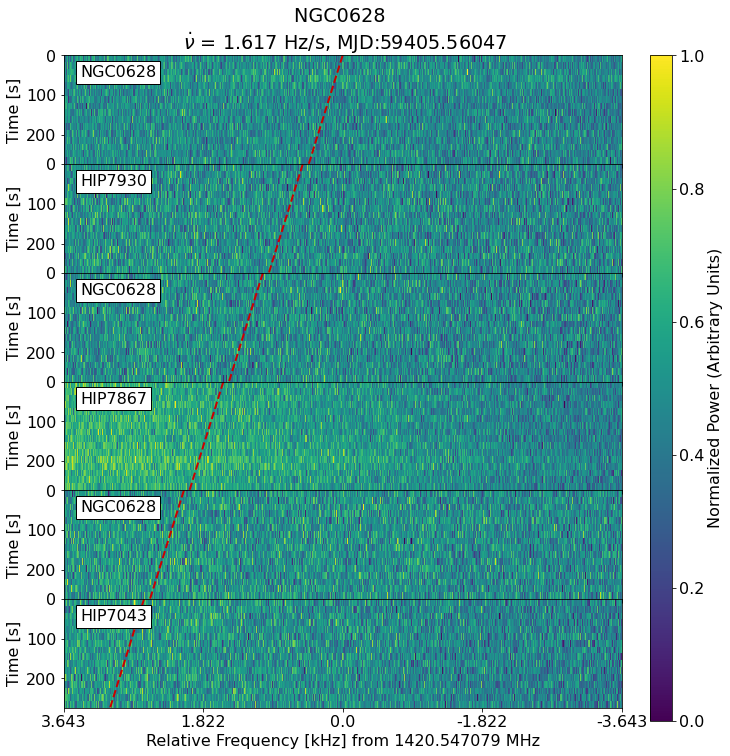}
    \caption{Full-cadence dynamic spectra for the two false-positive events found in the reduced \citet{Choza_2024} data set, with time from the beginning of the observation on the $y$-axis and frequency offset from the starting frequency of the event on the $x$-axis. Each block in a stack is one of the six pointings that make up the ABACAD observing cadence. The event from the DDO\,210 observation was also found by \citet{Choza_2024}, while the event from the NGC\,628 observation is new.}
    \label{fig:events}
\end{figure*}

PCA has been used as a preprocessing step in other clustering frameworks when a lower-dimensional \citep{Sooknunan:2021, Biswas:2023} and/or decorrelated \citep{Matchev:2022} feature space is needed.
However, the features found by PCA do not share equal importance. Instead, they contain different fractions of the total variance of the original data set, permitting the reconstruction of its broadest structures with only the few highest-variance features (``principal components"), but requiring more for finer details.

We used PCA to investigate our feature set's reducibility, since some correlation is present and some features are noisy.
Figure \ref{fig:PCA} shows the fraction of the total variance explained by increasing numbers of principal components. For some data sets, especially in the case where PCA is the only dimensionality reduction technique and is used to reduce the size of, e.g., a long time series from order $\gtrsim$10$^3$ to order $\sim$10, enough noise is present that only the first few principal components are necessary to capture all the salient information in the data; all the other components only serve to reconstruct the noise. However, for our already-reduced data set, virtually all the variance is significant. The only ``noise'' would come from features that are highly correlated with other features, or are otherwise somehow superfluous in capturing the statistics of the data.

The curve in Figure \ref{fig:PCA} is fairly smooth and does not jump suddenly from a low variance fraction to a fraction near 1.0. More than 99\% of the variance can be retained with as few as 10 components, but in a parameter space where much RFI looks very alike, even 1\% can be significant for clustering and anomaly detection. For lack of any obvious cutoff point, we keep all 13 original features and forgo PCA as a preprocessing step for our analysis. Using the full 13 principal components may also be a valid approach for reasons of decorrelation, since the principal components are linearly independent by definition, but we do not find any correlations in our feature space strong enough to merit PCA for their mitigation.

We do employ PCA to reduce the dimensionality of the centroid data set before applying the t-SNE algorithm, as discussed in Section \ref{match}.

We also used PCA to test the feasibility of anomaly ranking. Even after HDBSCAN reduction, the anomalies class may still require overwhelming manual follow-up for extremely large data sets, so some form of ranking scheme is desirable to guide the review. We ranked each anomaly by the Euclidean distance to its nearest non-anomalous neighbor, a metric intended to designate ``most anomalous" those signals that are farthest in parameter space from non-anomalous loci. The sensitivity of Euclidean distance decreases with dimensionality, so we project the anomalies into a lower-dimensional principal component space to give this metric more weight. In trials from 6 through 11 principal components kept, however, we found no significant improvement in the recovery of injected technosignature-like signals.

\section{Results}\label{results}

\begin{figure*}
    \centering
    \includegraphics[width=\textwidth]{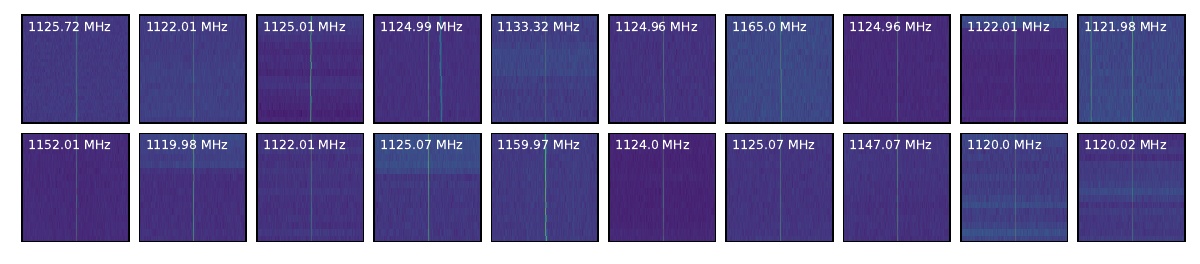}
    \includegraphics[width=\textwidth]{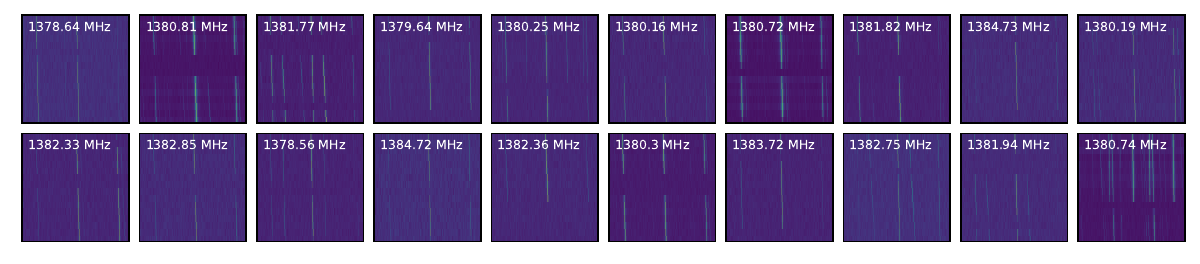}
    \includegraphics[width=\textwidth]{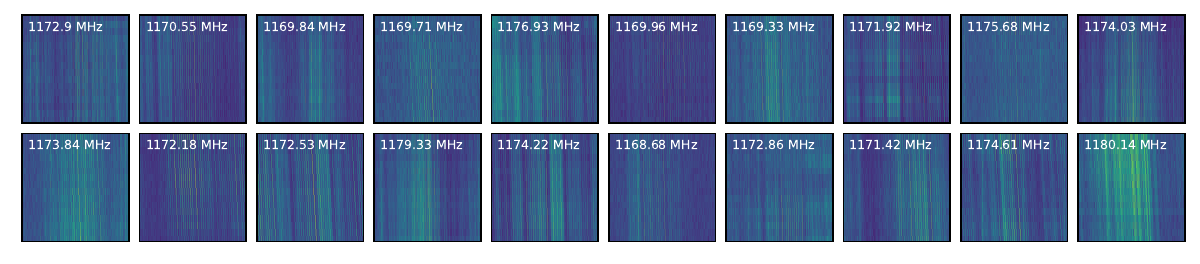}
    \includegraphics[width=\textwidth]{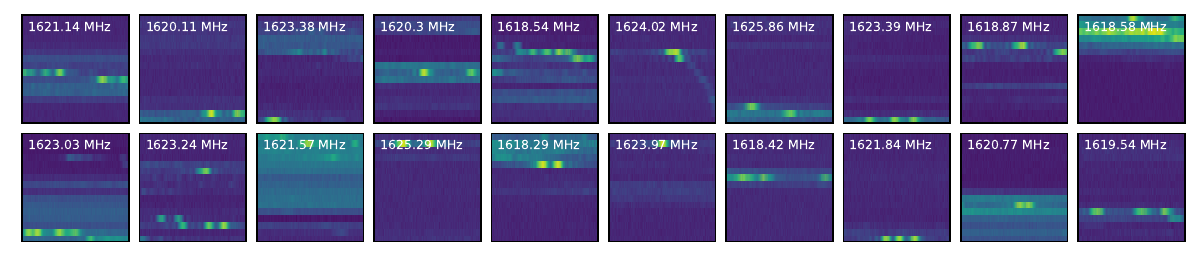}
    \caption{Examples of homogeneous clusters from batches before the cross-batch matching process, showing (top to bottom) low-drift narrowband signals consistent with aeronautical radionavigation, sparse comb structure consistent with GPS L3 signals near 1381 MHz, dense comb structure consistent with GPS L5 signals near 1176 MHz, and broad pulsed signals consistent with the Iridium satellite constellation between 1610 and 1626.5 MHz. These classes are not an all-inclusive portrait of the RFI environment of the \citet{Choza_2024} survey, but highlight generally common morphologies. The 20 samples shown for each cluster are selected at random. Frequency is on the $x$-axis and is 2.7\,kHz for all snippets. Time is on the $y$-axis and is 5\,min for all snippets.}
    \label{fig:clusters}
\end{figure*}

\begin{center}
\begin{table*}
\centering
\begin{threeparttable}
\setlength{\tabcolsep}{0pt}
\vspace{0.6cm}
    \caption{Effects of GLOBULAR clustering on the \tseti\ pipeline.}
    \begin{tabular*}{\textwidth}{@{\extracolsep{\fill}} l c c c c}
        \hline
        \hline
         & FP\tnote{a}~ Hits & FP Events & TP\tnote{b}~ Hits & TP Events \\
        \hline
        Without GLOBULAR clustering & 1,917,903 & 288 & 296 & 86 \\ 
        With GLOBULAR clustering & 132,885 & 2\tnote{c} & 264 & 69 \\
        \hline
        Reduction & 93.1\% & 99.3\% & 10.8\% & 19.8\% \\
        \hline
    \end{tabular*}
    \begin{tablenotes}
        \footnotesize
        \item[a] False positive; i.e., known RFI previously analyzed by \citet{Choza_2024}.
        \item[b] True positive; i.e., injected synthetic technosignatures.
        \item[c] Of the two false-positive events, one is a match to an event found by \citet{Choza_2024}. The other is new, though it is not a compelling technosignature candidate.
    \end{tablenotes}
    \label{tab:comparison_table}
\end{threeparttable}
\end{table*}
\end{center}

On the $L$-band portion of the \citet{Choza_2024} data set, we obtain a 93.1\% reduction in the false-positive hit rate after 8 epochs, and a 99.3\% reduction in the corresponding event rate (from 288 to 2). The data set was processed in batches of approximately 3,000 hits each, plus 100 synthetic ``seed'' signals that were the same for all batches. HDBSCAN was applied to each batch with the same hyperparameters: $n_\text{pts}=$ 4, $\rho_\text{pts}=$ 2, and $\epsilon_m=$ 0.18.

Table \ref{tab:comparison_table} summarizes our results. Of the two events discovered by the \verb+FindEvent+ pipeline on our processed hit data set, one is a clear match to an event found by \citet{Choza_2024}, while one is new. Both are shown in Figure \ref{fig:events}; by visual inspection, neither is a compelling technosignature candidate. Curiously, the new event, in an observation of NGC\,628, was found in the Galactic 21-cm line near 1420 MHz and appears to correspond to stochastic intensity variations rather than an actual narrowband signal, anthropogenic or otherwise. 
We speculate that the presence of the 21-cm line may bias the calculation by \verb+turboSETI+ of the rms noise across the 3-MHz coarse channel that contains it, allowing intensity spikes much lower than the SNR threshold of the search to be elevated above the threshold and found erroneously as hits. Using the \verb+FindDoppler+ pipeline, we find numerous other hits in the region that similarly do not appear to correspond to actual signals. The finding of the event near the 21-cm line is noteworthy in part because of the hit spectral occupancy of this region. It suggests that RFI rejection in RFI-dense spectral regions can help the \verb+FindEvent+ pipeline find events it would otherwise miss. We defer an analysis of the requisite RFI densities, and consequently the cases in which this new event finding may be practicable, to future work.

We also tested our method's recovery rate of injected synthetic signals. We simulated 100 drifting narrowband signals with drift rates between $-3$ and $3$\,Hz\,s$^{-1}$ and SNRs between 1,000 and 10,000, and injected them into the ``on'' pointings of cadence observations of M\,31, a total of 300 hit injections. This is a different set of synthetic signal injections than the 100 seed signals used to identify clusters of interest in each batch. Of the 300 hit injections, 296 were identified by the \verb+FindDoppler+ hit search, corresponding to a possible 98 out of the 100 injected events. From the identified injections, the GLOBULAR clustering process erroneously removed 32 hits, leaving 264.

It is known that, due to a ranking algorithm that finds the ``top hit'' in each small frequency window, defined by the maximum drift rate of the search, \verb+turboSETI+ is insensitive to multiple hits occurring close together in frequency \citep{Margot-2021}. This appears to account for two of the four signals missed by the \verb+FindDoppler+ search, but does not explain why the \verb+FindEvent+ search fails to link some of the correctly identified signals across their respective cadences.

After applying the spatial filter, 86 of the 100 synthetic events were recovered by \verb+turboSETI+ alone and 69 were recovered with GLOBULAR clustering. All of those recovered with GLOBULAR clustering were also recovered by \verb+turboSETI+ alone. Typically, an event missed by GLOBULAR clustering was due to a hit being removed from a single ``on'' scan, which disqualifies the signal from event status under the high-duty-cycle assumption. We use \verb+FindEvent+ for consistency with \citeauthor{Choza_2024} and other BL searches, but refinements in the spatial filter may enable the recovery of further true positives.
Additionally, since 98 of the 100 injected events exist in the post-\verb+FindDoppler+ data set, the ceiling for potentially recoverable events is likely higher than the 86 found by the \verb+turboSETI+-alone pipeline. The discovery of a new false-positive event not found by the \verb+turboSETI+-alone analysis suggests that GLOBULAR clustering may be able to help the \verb+turboSETI+ algorithm find some of the true-positive events it previously missed. We defer this investigation to future work.

To be assured of valid clustering, we manually inspected a random sample of single-batch clusters, of which four examples are shown in Figure \ref{fig:clusters}, and identified likely sources. Finally, on a sample of 10,000 hits clustered out as RFI, we performed the cross-batch cluster matching scheme outlined in Section \ref{match} and inspected the aggregate clusters. As expected, the single-batch clusters were more homogeneous on average than the cross-batch clusters, though many of the cross-batch clusters do display comparable homogeneity to single-batch clusters. Most of the inhomogeneous clusters are clear composites of two or three homogeneous clusters, a refinable error in the application of HDBSCAN to the t-SNE embedding, and the remaining inhomogeneities present a good use case for refinement with active learning. We show examples of relatively homogeneous cross-batch clusters in Appendix \ref{appA}.

\section{Conclusions}

In this work, we demonstrate an application of the HDBSCAN clustering algorithm to technosignature searches, where it significantly reduces the false positive rate while maintaining an injected signal recovery rate only slightly lower than current search methods. HDBSCAN does not scale easily to data sets of the volume regularly seen in radio astronomy, but we mitigate the deleterious effects of upscaling with an iterative batching and pooling scheme. This implementation significantly reduces both noise and compute time over an approach that runs HDBSCAN once on all data points.

GLOBULAR clustering shows promise as a broad-brush RFI rejection algorithm. Its primary benefit is that it is agnostic by design to the types of RFI present in a data set. Unlike a convolutional neural network or other supervised learning method, which requires that the user prescribe RFI types for identification, GLOBULAR clustering finds all the common RFI types in a data set purely on the basis of their prevalence. Since RFI sources are extremely diverse and growing more so over time, an agnostic rejection framework is critical for robust, at-scale mitigation.

An additional benefit is that, by seeding new clusters with injected synthetic signals, GLOBULAR clustering can be optimized to search for specific signal types even if their morphologies are not unusual enough to intrinsically separate them as outliers. A theoretically arbitrary number of signal types can be specified like this, boosting the algorithm's utility for a wide range of survey targets. In the absence of signal seeding, truly anomalous signals are still in principle likely to be identified by being placed in the anomalous class through all epochs. Future work will quantify the types of true anomalies to which our method is most, and least, sensitive.

In our cross-batch cluster matching method, limitations remain. As discussed in Section \ref{match}, apparent clusters in t-SNE embeddings should be interpreted with caution. Manual inspection of our cross-batch clusters shows that many of them are comparably homogeneous to the single-batch clusters shown in Figure \ref{fig:clusters}, but not all. Future work may reduce the inhomogeneities, improving the ability of GLOBULAR clustering to precisely characterize an RFI environment. For example, refining the feature space, either by introducing new features or by eliminating insufficiently informative ones, may enable more reliable discrimination between certain easily confused RFI types. The exact makeup of the optimal feature space may depend on the instrument, with features that informatively characterize the GBT environment at $L$-band being less important at $X$-band, or entirely superfluous for another observatory. Iterative user experimentation, especially on small representative toy data sets, can be valuable for developing the optimal feature space, as discussed in Section \ref{toy}.

Active learning presents another potentially useful tool for refinement. Through a front-end interface, a user can manually relabel an active learning model's first guess at signal classifications, one signal at a time, to improve future classifications. By preferentially selecting signals that occur at the boundaries of clusters, the user need only relabel a small fraction of the data set for their feedback to dramatically improve the model's classification overall. This method has already been applied to great effect by \citet{Lochner_2021} in their package {\sc Astronomaly}, which was originally designed for anomaly detection in galaxy images but is extendable to other survey types and data products, including in the radio, as demonstrated by \citet{Andersson_2024}. As \citet{Etsebeth_2024} show, as few as $\sim 10^5$ user labels are necessary for effective classification and anomaly detection in a data set of 3,884,404 galaxy images, easily achievable by one person in a few hours. Similar principles are likely to apply for dynamic spectra in comparably sized radio surveys, and will be investigated in detail in future work.

Future refinements aside, GLOBULAR clustering as presented in this work brings significant advantages to a pipeline that includes it. The 99\% reduction of the false-positive event rate alleviates the manual follow-up requirements inherent to large surveys, which are otherwise likely to grow prohibitive with increasing data volumes. Moreover, the injected signal recovery rate is within 20\% of the recovery rate of \verb+turboSETI+ alone. The signal recovery rate is aided by our use of seed signals (Section \ref{batch}), and can be improved for any use case, not just drifting narrowband signal searches, by a more comprehensive seed signal set. In this way, GLOBULAR clustering is straightforwardly extendable both to future narrowband technosignature searches, such as those enabled by the development of the Breakthrough Listen Interesting Signal Search (BLISS) by \citet{West:BLISS}, and to other types of radio surveys.

\section*{Acknowledgements}

 The Breakthrough Prize Foundation funds the Breakthrough Initiatives, which manages Breakthrough Listen. The Green Bank Observatory facility is supported by the National Science Foundation, and is operated by Associated Universities, Inc., under a cooperative agreement. CP was funded as a participant in the Berkeley SETI Research Center Research Experience for Undergraduates Site, supported by the National Science Foundation under Grant No.~2244242. The authors thank the anonymous referee for valuable comments and Verlon Etsebeth and Peter Ma for helpful discussions, and BJB is grateful to the 2024 Berkeley SETI REU participants for their support.

 \software{NumPy \citep{harris2020array},
Matplotlib \citep{matplotlib},
Astropy \citep{astropy:2013, astropy:2018, astropy:2022},
SciPy \citep{SciPy},
Pandas \citep{mckinney_pandas_2010},
h5py \citep{collette_python_hdf5_2014},
Scikit-Learn \citep{scikit-learn},
HDBSCAN \citep{McInnes2017},
Blimpy \citep{Price_blimpy},
turboSETI \citep{turboSETI},
setigen \citep{Brzycki_2022},
SHAP \citep{shap}
}

 \appendix

 \section{Cross-Batch Hit Clusters}\label{appA}

\begin{figure*}[!hpt]
    \centering
     \includegraphics[width=\textwidth]{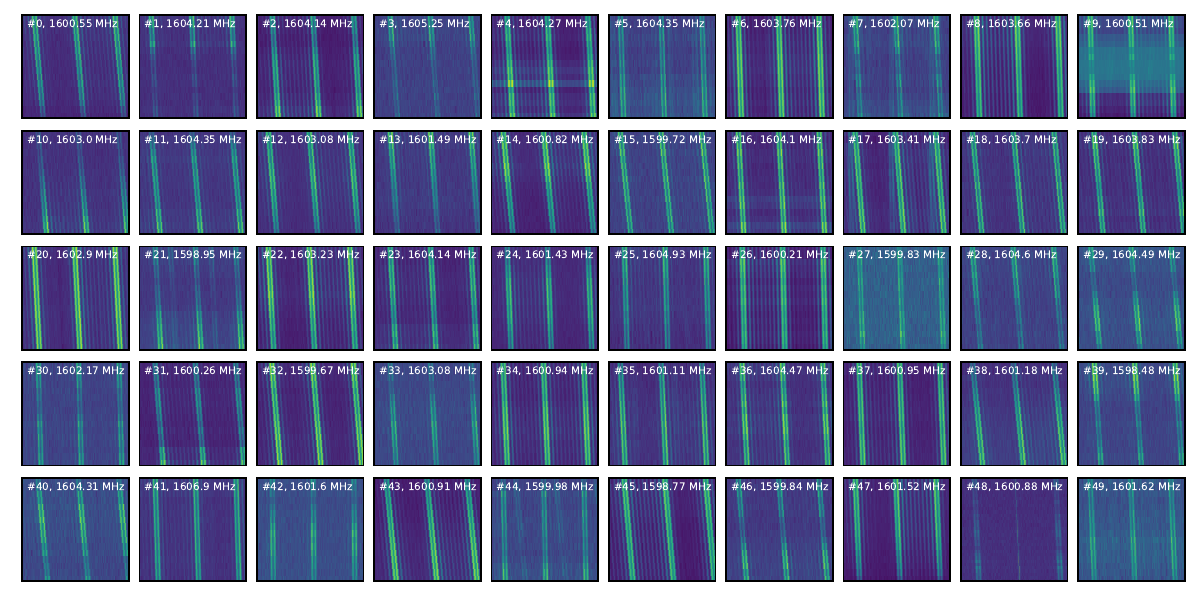}
     \caption{Snippets of dynamic spectra for 50 randomly selected hits from Cluster 0 in the t-SNE/HDBSCAN post-batching RFI analysis (see Figure \ref{fig:tSNE}).}
     \label{fig:class0}
 \end{figure*}

\begin{figure*}[!hpt]
     \centering
     \includegraphics[width=\textwidth]{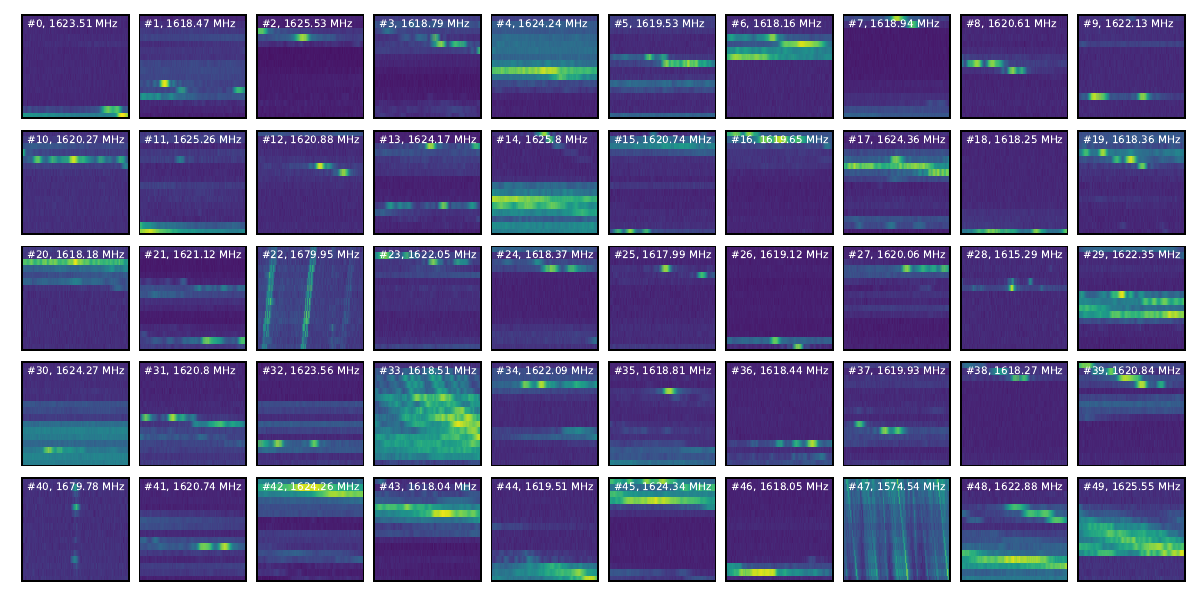}
     \caption{Snippets of dynamic spectra for 50 randomly selected hits from Cluster 17 in the t-SNE/HDBSCAN post-batching RFI analysis (see Figure \ref{fig:tSNE}).}
     \label{fig:class17}
\end{figure*}

\begin{figure*}[!hpt]
     \centering
     \includegraphics[width=\textwidth]{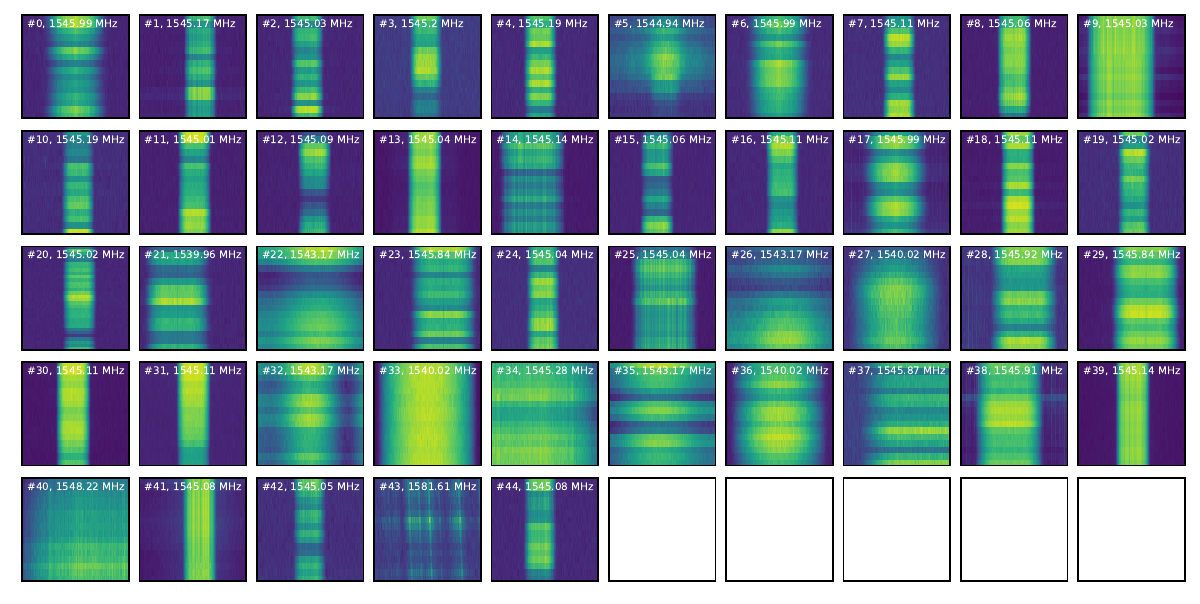}
     \caption{Snippets of dynamic spectra for the 45 hits from Cluster 19 in the t-SNE/HDBSCAN post-batching RFI analysis (see Figure \ref{fig:tSNE}).}
     \label{fig:class19}
\end{figure*}

\begin{figure*}[!hpt]
     \centering
     \includegraphics[width=\textwidth]{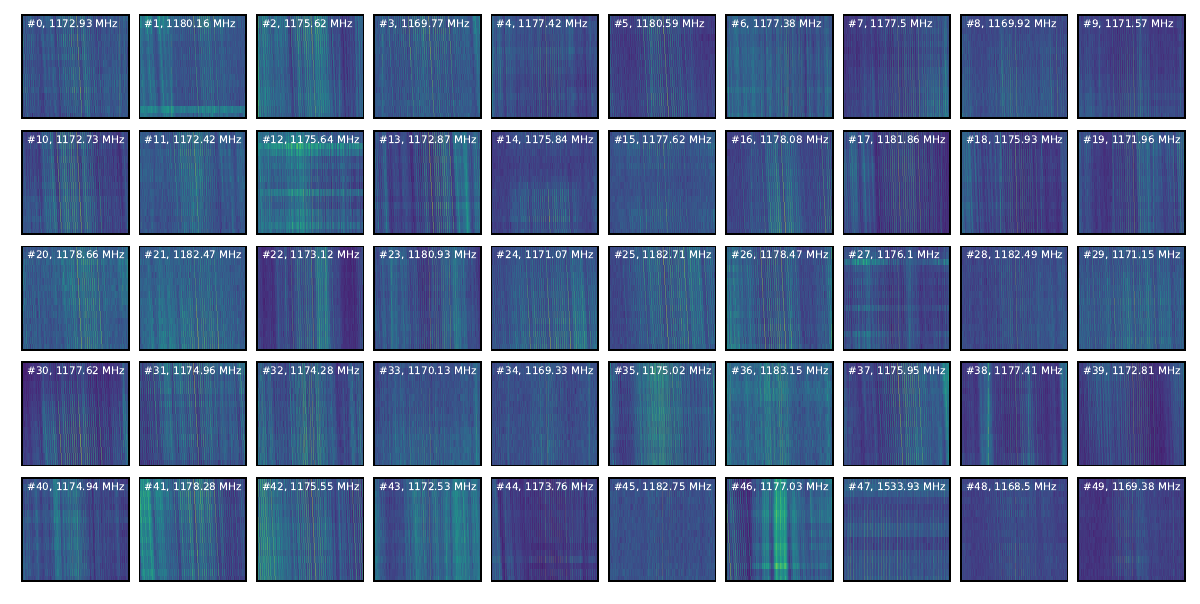}
     \caption{Snippets of dynamic spectra for 50 randomly selected hits from Cluster 21 in the t-SNE/HDBSCAN post-batching RFI analysis (see Figure \ref{fig:tSNE}).}
     \label{fig:class21}
\end{figure*}

\begin{figure*}[!hpt]
     \centering
     \includegraphics[width=\textwidth]{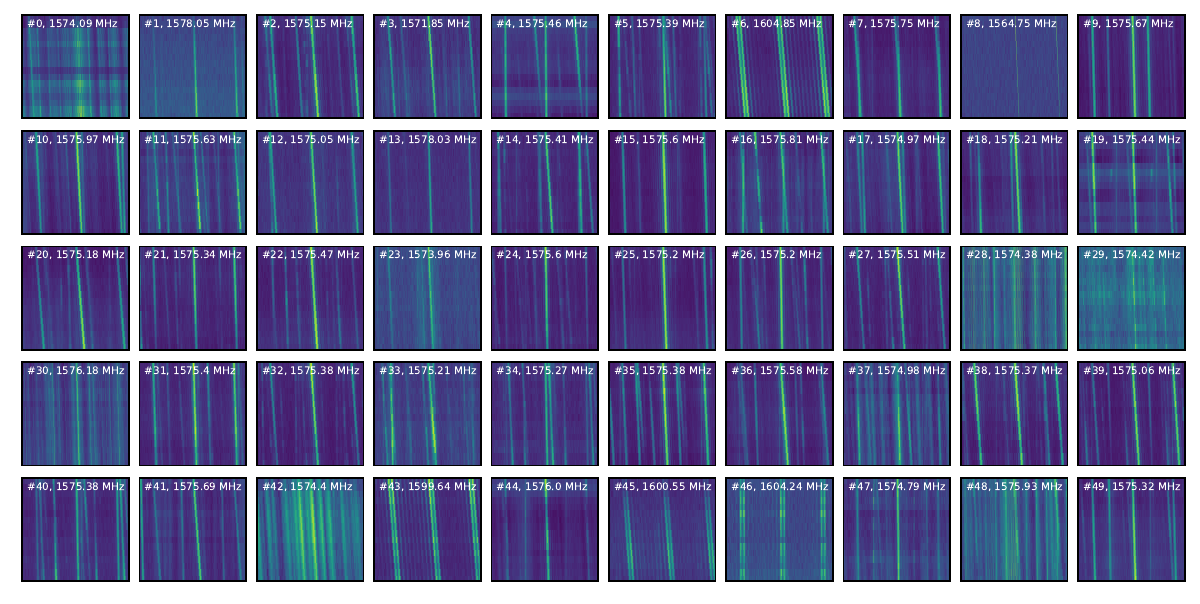}
     \caption{Snippets of dynamic spectra for 50 randomly selected hits from Cluster 27 in the t-SNE/HDBSCAN post-batching RFI analysis (see Figure \ref{fig:tSNE}).}
     \label{fig:class27}
\end{figure*}

 Figures \ref{fig:class0}–\ref{fig:class27} show samples of hits from relatively homogeneous clusters that have been matched across batches. As shown in Figure \ref{fig:tSNE}, there are 58 clusters in total, not counting outliers, though some clusters correspond to morphologically similar RFI. The hits in each cluster are selected at random (except for Cluster 19, for which all hits are shown) and sorted by distance from the cluster's centroid in the t-SNE embedding space, with the nearest first. Frequency is on the $x$-axis and is 2.7\,kHz for all plots, while time is on the $y$-axis and is 5\,min for all plots.


 \clearpage

\bibliographystyle{aasjournal}
\bibliography{main.bib}

\end{document}